\documentclass[conference,twocolumn,10pt]{IEEEtran}

\pdfoutput=1
\usepackage[cmex10]{amsmath}
\usepackage{amsfonts,amssymb}
\usepackage{verbatim}
\usepackage{color}
\usepackage{multirow} 

\usepackage{cite}

\ifCLASSINFOpdf
  \usepackage[pdftex]{graphicx}
  \graphicspath{{graphics/}}
  \DeclareGraphicsExtensions{.pdf,.jpeg,.png}
\else
  \usepackage[dvips]{graphicx}
  \graphicspath{{graphics/}}
  \DeclareGraphicsExtensions{.eps}
\fi

\usepackage{array}

\usepackage{mdwmath}
\usepackage{mdwtab}
\begin{document}
\bibliographystyle{IEEEtran}
%
\title{Multi-Scale Stochastic Simulation for Diffusive Molecular Communication}

\author{\IEEEauthorblockN{Adam Noel$^{\ast}$, Karen C.
Cheung$^{\ast}$, and Robert Schober$^{\ast\dagger}$}
\IEEEauthorblockA{$^{\ast}$Department of Electrical and Computer
Engineering\\
University of British Columbia, Email: \{adamn, kcheung, rschober\}@ece.ubc.ca
\\ $^{\dagger}$Institute for Digital Communications\\
Friedrich-Alexander-Universit\"{a}t Erlangen-N\"{u}rnberg (FAU), Email:
schober@LNT.de}}


\newcommand{\x}{x}
\newcommand{\y}{y}
\newcommand{\z}{z}


\newcommand{\kth}[1]{k_{#1}}

\newcommand{\M}{M}
\newcommand{\A}{A}
\newcommand{\Dx}[1]{D_{#1}}
\newcommand{\Nx}[1]{N_{#1}}

\newcommand{\EXP}[1]{\exp\left(#1\right)}


\newcommand{\Vol}[1]{V_{#1}}
\newcommand{\subV}[1]{S_{#1}}
\newcommand{\dt}[1]{t_{#1}}
\newcommand{\prop}[1]{a_{#1}}
\newcommand{\molNum}[1]{U_{#1}}

\newcommand{\new}[1]{\textbf{#1}}

\newcommand{\edit}[2]{#1}

\maketitle

\begin{abstract}
Recently, hybrid models have emerged that combine microscopic and mesoscopic regimes in a single stochastic reaction-diffusion simulation. Microscopic simulations track every individual molecule and are generally more accurate. Mesoscopic simulations partition the environment into subvolumes, track when molecules move between adjacent subvolumes, and are generally more computationally efficient. In this paper, we present the foundation of a multi-scale stochastic simulator from the perspective of molecular communication, for both mesoscopic and hybrid models, where we emphasize simulation accuracy at the receiver and efficiency in regions that are far from the communication link. Our multi-scale models use subvolumes of different sizes, between which we derive the diffusion event transition rate. Simulation results compare the accuracy and efficiency of traditional approaches with that of a regular hybrid method and with those of our proposed multi-scale methods.
\end{abstract}

\section{Introduction}

Molecular communication (MC), which is the use of molecules as information carriers to carry data from a transmitter to an intended receiver, is ubiquitous for signalling in biological systems; see \cite[Ch.~16]{RefWorks:588}. Interest has recently emerged in adapting the principles of MC to deliver arbitrary amounts of information in synthetic communication networks, with the goal of enabling new applications in areas that include biological engineering and manufacturing; see \cite{RefWorks:801}. The simplest method of MC, free diffusion, is a passive strategy where molecules collide with other molecules while propagating from the transmitter to the receiver. Diffusion requires a fluid medium from the source to the destination but no other external infrastructure. It is also very fast over short distances (e.g., bacterial cells rely on diffusion for their internal signaling requirements), but is an inherently noisy process due to the randomness of molecular collisions.

Meaningful communications analysis requires a detailed understanding of the propagation environment. Generally, we desire the form and the statistics of the end-to-end channel impulse response, i.e., the time-varying signal observed at the receiver given that molecules are released at some instant by the transmitter, and the response can then be used to derive the received signal for any modulation scheme. However, even the \emph{expected} channel impulse response of diffusive MC cannot be analytically derived in closed form, except with simplifying assumptions and specific system geometries; see \cite{RefWorks:586}. For example, we analyzed an unbounded environment with diffusion, bulk fluid flow, and molecule degradation in \cite{RefWorks:747}. The examination of more complex and more realistic environments generally requires the use of numerical methods or simulations to generate the end-to-end statistics. For example, the expected channel impulse response was derived in the Laplace domain for a bounded diffusive environment and various receiver reaction pathways in \cite{RefWorks:829}.

Reaction-diffusion simulators make a tradeoff between simulation accuracy and computational efficiency. The most accurate simulators are termed \emph{microscopic} and they track the coordinates of every individual molecule in the system, e.g., the Smoldyn simulator in \cite{RefWorks:622}. Microscopic simulators are arguably the most computationally expensive, especially as the total number of molecules in the system increases and if the molecules can participate in many types of reactions.

\emph{Mesoscopic} simulators partition the system into subvolumes or containers and are generally more computationally efficient as system complexity increases. Reactions involving two or more molecules can only occur if the reactants are all in the same subvolume, and diffusion is represented as a reaction that transfers a molecule between two subvolumes. If the molecular populations throughout each subvolume are homogeneous, i.e., \emph{well-stirred}, and the size of each subvolume is appropriate for the time scale of the reactions that can occur, then mesoscopic simulations that execute one reaction at a time can accurately capture system behavior; see \cite{RefWorks:617}.

Mesoscopic simulators have three characteristics that can make them more computationally efficient than microscopic simulators. First, the required memory grows with the number of subvolumes and the number of possible reactions within a subvolume; the memory required in a microscopic model grows with the total number of molecules. Second, a mesoscopic model can easily account for the execution of arbitrarily complex chemical reaction pathways (assuming that a subvolume is well-stirred), whereas accounting for reactions with more than one reactant in a microscopic model is a combinatorial problem where we must compare the relative positions of every combination of potential reaction participants; see \cite{RefWorks:623}. Third, if the molecular populations within subvolumes are sufficiently large, then mesoscopic models can execute multiple reactions simultaneously without a significant loss in temporal accuracy; see \cite{RefWorks:612}.

For communications analysis, we are ultimately interested in accurately modeling what happens at the receiver. We do not need to model the detailed evolution of the entire system; a computationally efficient evolution of the overall system is sufficient \emph{if} we can maintain accuracy at the receiver. For simple systems, e.g., diffusion only, accurate microscopic simulations can be obtained without excessive computational resources, so we might be willing to simulate accurately ``everywhere''. However, microscopic simulations with complex reaction kinetics and large molecular populations (with more than hundreds of thousands of molecules) become computationally prohibitive, especially if we need to simulate thousands of realizations in order to generate the desired statistics. One solution is to simplify the kinetics, as we have done in our previous work in \cite{RefWorks:747}. Another alternative is to use a mesoscopic simulator, but we might lose the desired accuracy at the receiver.

Recent advances in stochastic reaction-diffusion models have provided new methods to increase computational efficiency without a significant loss in accuracy throughout the system of interest. The size of subvolumes in a mesoscopic model was adjusted in both space and time in \cite{RefWorks:806}. Larger subvolumes are more efficient (because there are fewer total subvolumes \emph{and} fewer diffusion transitions between them), but generally less accurate because a larger area must be well-stirred. Hybrid strategies that combine microscopic and mesoscopic models have been proposed in \cite{RefWorks:870,RefWorks:869,RefWorks:871,RefWorks:872}.

Simulation frameworks that are publicly available and designed specifically for MC research, such as BiNS2 \cite{RefWorks:736} and MUCIN \cite{RefWorks:890}, are generally based on microscopic models. We claim that there is no computationally efficient end-to-end simulation framework in use by the wider MC community that can accommodate complex reaction pathways. One barrier is the fundamental tradeoff of accuracy and efficiency. In this paper, we present the foundation of a multi-scale stochastic simulator. We integrate recent advances in stochastic reaction-diffusion models to maximize accuracy at the receiver while emphasizing efficiency elsewhere. Specifically, we combine subvolumes of different sizes with a hybrid model. Subvolumes in the mesoscopic regime that are far from the transmitter and receiver can be made larger to improve efficiency. The primary contributions of this paper are summarized as follows:
\begin{enumerate}
    \item We derive the transition rate between subvolumes whose adjacent faces do not completely overlap (conventionally, adjacent subvolumes share a common face). An expression for the transition rate was derived in \cite{RefWorks:806}. Our derivation is geometrically simpler, more intuitive, and is consistent in that conservation laws are satisfied (i.e., there is no net ``leaking'' of molecules to or from larger subvolumes in a uniform environment).
    \item We present a novel hybrid micro/meso simulation algorithm. For clarity of presentation, we adopt the basic rules described in \cite{RefWorks:870} to determine the transitions between the microscopic and mesoscopic regimes.
    \item We compare the performance of conventional simulation models and a regular hybrid model with our proposed multi-scale models for two simulation test cases.
\end{enumerate}

The treatment of transitions between microscopic and mesoscopic regimes and between subvolumes of different sizes does not affect chemical reactions because they can only occur within the microscopic regime or within a single subvolume. In addition, the transition between regimes of three dimensions is an analogous extension of the transition between two-dimensional regimes, as shown in \cite{RefWorks:869}. Thus, we focus on two-dimensional diffusion for clarity of presentation and to not place too many constraints on subvolume size. We note that the computational benefits of mesoscopic models become more \edit{pronounced}{} in more complex environments, which will be considered in our future work.
    
The rest of this paper is organized as follows. We define the physical environment and review existing simulation models in Section~\ref{sec_model}. In Section~\ref{sec_diff}, we derive the transition rate between square subvolumes whose adjacent faces do not completely overlap, present rules for transitioning between the microscopic and mesoscopic regimes, and describe our hybrid micro/meso algorithm. The simulation frameworks are compared in Section~\ref{sec_results}. Conclusions are drawn in Section~\ref{sec_concl}.

\section{System Model and Preliminaries}
\label{sec_model}

\begin{figure}[!tb]
	\centering
	\def\svgwidth{\linewidth}
	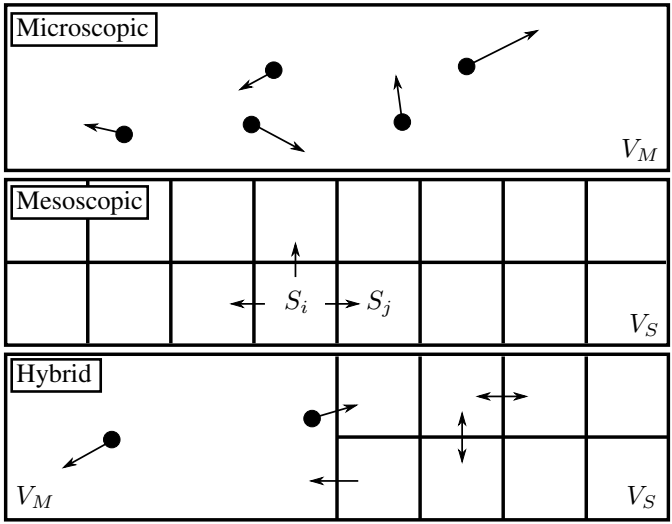
	\caption{Comparison of three classes of stochastic diffusion models. The microscopic model (top) tracks each molecule (black circle) and its precise coordinates in $\Vol{\M}$. The mesoscopic model (middle) partitions the environment $\Vol{\subV{}}$ into subvolumes and molecules can move between adjacent subvolumes, shown here as squares. The hybrid model (bottom) includes both regimes, such that molecules can transition between $\Vol{\M}$ and $\Vol{\subV{}}$.}
	\label{fig_model_comp}
\end{figure}

In this section, we describe the diffusive environment and then discuss the existing approaches for simulating diffusion (as summarized in Fig.~\ref{fig_model_comp}). Our discussion is by no means exhaustive, as many variations of microscopic and mesoscopic simulators exist, but we provide sufficient detail to describe our implementation of multi-scale diffusion simulations.

We model a two-dimensional fluid environment $\Vol{}$ that could be bounded or unbounded. We will assume that the boundary of $\Vol{}$ is reflective; \edit{generally, it could be absorbing or reactive if chemical reactions are introduced at the boundary}{}. There is a single molecular species with constant diffusion coefficient $\Dx{}$, as we considered in \cite{RefWorks:747}.

\subsection{Microscopic Model}

In the microscopic model, the environment is effectively a single large container $\Vol{\M}$. The precise coordinates of every molecule in $\Vol{\M}$ are tracked. If the cartesian coordinates of the $i$th molecule at time $t$ are $\{\x_i(t),\y_i(t)\}$, then the coordinates of that molecule at time $t + \Delta t_{\M}$ are updated as \cite[Eq.~(1.7)]{RefWorks:869}
\begin{align}
\label{eq_diff_x}
\x_i(t + \Delta t_{\M}) = &\x_i(t) + n_1\sqrt{2\Dx{}\Delta t_{\M}},\\
\label{eq_diff_y}
\y_i(t + \Delta t_{\M}) = &\y_i(t) + n_2\sqrt{2\Dx{}\Delta t_{\M}},
\end{align}
where $n_1$ and $n_2$ are independent normal random values with mean 0 and variance 1. Candidate destinations that are outside of $\Vol{\M}$ are reflected against the boundary of $\Vol{\M}$ as necessary.

\subsection{Mesoscopic Model}

In the mesoscopic model, the environment $\Vol{\subV{}}$ is partitioned into non-overlapping subvolumes. \edit{Limits on subvolume size include being small enough to remain well-stirred and large enough for bimolecular reactants to be in the same subvolume if they are close enough to react; see \cite{RefWorks:613,RefWorks:617}}{}. We track the total number of molecules inside each subvolume, and we assume that the molecules within a given subvolume are uniformly distributed. A diffusion \emph{event} occurs when a molecule jumps between two adjacent subvolumes, e.g. leaving subvolume $\subV{i}$ and entering subvolume $\subV{j}$ (see Fig.~\ref{fig_model_comp}).

Reactions in mesoscopic models, including diffusion ``reactions'', are simulated by assigning reaction \emph{propensities} to every possible reaction ``event''; see \cite{RefWorks:616}. We define the propensity for a molecule to diffuse from subvolume $\subV{i}$ to $\subV{j}$ as $\prop{i,j}$. The probability of this diffusion event occurring within an infinitesimal time step $\delta t$ is then $\prop{i,j}\delta t$. If these two subvolumes are adjacent squares of width $h$, then the propensity is given by \cite[Eq.~(1.6)]{RefWorks:869}
\begin{equation}
\label{eq_diff_prop}
\prop{i,j} = \frac{\Dx{}}{h^2}\molNum{i},
\end{equation}
where $\molNum{i}$ is the number of $\A$ molecules inside $\subV{i}$. We derive the diffusion propensity between subvolumes whose faces do not fully overlap in Section~\ref{sec_diff}. If there is no overlap at all, i.e., if the subvolumes $\subV{i}$ and $\subV{j}$ are not neighbors, then $\prop{i,j}=0$.

There are several (mathematically equivalent) methods for using the propensities to simulate a sequence of reaction events and the times when they occur. Here, we consider a variation of the \emph{direct method}, which is arguably the simplest to implement; a more complete discussion of the motivation for this method and its alternatives can be found in \cite{RefWorks:620}. First, we determine the total propensity $\prop{i}$ for each subvolume, i.e.,
\begin{equation}
\label{eq_sub_prop}
\prop{i} = \sum_j \prop{i,j},
\end{equation}
and we find the total system propensity $\prop{T}=\sum_i \prop{i}$. Next, we generate three independent uniform random numbers between 0 and 1: $u_1$, $u_2$, and $u_3$. The time $\dt{\subV{}}$ of the next diffusion event after time $t$ can be realized by evaluating \cite[Eq.~(1)]{RefWorks:620}
\begin{equation}
\label{eq_meso_time}
\dt{\subV{}} = t-\left(\log{u_1}\right)/\prop{T}.
\end{equation}

\edit{We determine the subvolume $\subV{i}$ where the diffusion event originates by performing a weighted die roll, i.e., subvolume $\subV{q}$ has a probability of $\prop{q}/\prop{T}$ of being chosen. We perform the die roll by comparing the value of $u_2$ with the cumulative sum of $\prop{q}/\prop{T}$ terms.}{}


\edit{Finally, we determine the specific destination subvolume $\subV{j}$ using another weighted die roll, i.e., subvolume $\subV{r}$ has a probability of $\prop{i,r}/\prop{i}$ of being chosen. We perform the die roll by comparing the value of $u_3$ with the cumulative sum of $\prop{i,r}/\prop{i}$ terms.}{}
For example, determining $\subV{j}$ for subvolume $\subV{i}$ in Fig.~\ref{fig_model_comp} is equivalent to drawing a random number from $\{1,2,3\}$ with equal probability. Given $\subV{j}$, we execute the diffusion event by decrementing $\molNum{i}$ and incrementing $\molNum{j}$. We update all of the propensities of $\subV{i}$ and $\subV{j}$ via (\ref{eq_diff_prop}), update $\prop{i}$ and $\prop{j}$ via (\ref{eq_sub_prop}), and continue the event sequence with three new uniform random numbers and find the time until the next diffusion event via (\ref{eq_meso_time}).

\subsection{Hybrid Model}

In the hybrid model, the environment $\Vol{}$ is partitioned into two regimes: a microscopic regime $\Vol{\M}$ and a mesoscopic regime $\Vol{\subV{}}$, such that $\Vol{}$ is the union of the closures of $\Vol{\M}$ and $\Vol{\subV{}}$. Each regime is treated independently, except when there are transitions across the interface between the two regimes. We discuss the details of these transitions in Section~\ref{sec_hybrid_transition}.

\section{Multi-Scale Diffusion}
\label{sec_diff}

In this section, we describe the key components of our multi-scale stochastic diffusion simulator. We derive the transition rate between square subvolumes with adjacent faces that do not completely overlap. We describe our transition rules to handle molecules that move from the microscopic regime to the mesoscopic regime and vice versa. Finally, we summarize our hybrid micro/meso simulation algorithm, which is based in part on the two-regime method described in \cite{RefWorks:869}.

\subsection{Diffusion Between Subvolumes of Different Size}

\begin{figure}[!tb]
	\centering
	\def\svgwidth{\linewidth}
	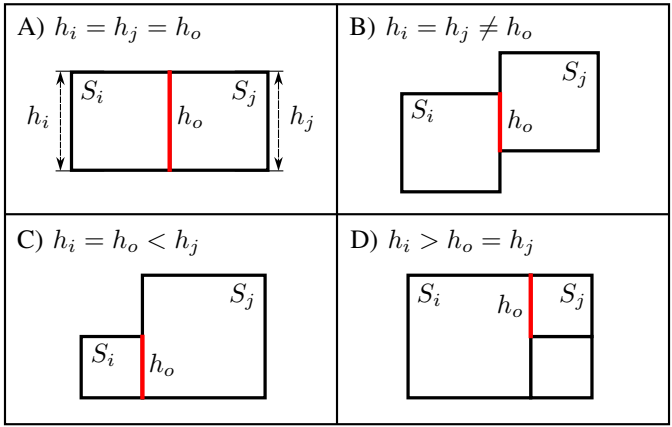
	\caption{Examples of overlap between adjacent subvolumes $\subV{i}$ and $\subV{j}$ with lengths $h_i$ and $h_j$, respectively. $h_i$ and $h_j$ are only shown explicitly in example A). In each example, the length of the overlap $h_o$ is labeled in red. Examples A) and B) have subvolumes of the same size. Examples C) and D) have subvolumes of different sizes. \edit{A case like example B), where $h_o$ is not equal to $h_i$ or $h_j$, might occur when accounting for the shape of the boundary of $\Vol{}$, or if $h_i \ne h_j$ and neither is a multiple of the other.}{}}
	\label{fig_sub_different}
\end{figure}

Consider two adjacent squares, of width $h_i$ and $h_j$, such that the overlap of the adjacent faces is of length $h_o \le \min\{h_i,h_j\}$. Examples are shown in Fig.~\ref{fig_sub_different}. We seek an expression for the diffusion propensity $\prop{i,j}$ from $\subV{i}$ to $\subV{j}$, which has the general form $\prop{i,j} = \kth{i,j}\molNum{i}$ (see \cite{RefWorks:613}), so we must derive the transition rate $\kth{i,j}$. We begin with a one-dimensional approximation of diffusion (a more general approach, such as that in \cite{RefWorks:806}, results in a propensity that can be a function of the number of molecules in a common neighbor subvolume $\subV{n}$). The transition rate for one-dimensional subvolumes of lengths $h_i$ and $h_j$ has been derived as \cite[Eq.~(15)]{RefWorks:613}
\begin{equation}
\label{eq_diff_rate_1d}
\kth{i,j} = \frac{2\Dx{}}{h_i(h_i + h_j)}.
\end{equation}

Eq.~(\ref{eq_diff_rate_1d}) is valid in more than one dimension \emph{if} the adjacent face of $\subV{i}$ is fully overlapped by that of $\subV{j}$. Equivalently, (\ref{eq_diff_rate_1d}) is valid if a molecule leaves $\subV{i}$ in the direction of $\subV{j}$ and the only possible destination in that direction is $\subV{j}$, i.e., as in examples A) or C) in Fig.~\ref{fig_sub_different}. Eq.~(\ref{eq_diff_rate_1d}) is not true for every overlap length $h_o$ between two adjacent squares, even if $h_i=h_j$ (as in example B) in Fig.~\ref{fig_sub_different}). We account for $h_o$ by scaling $\kth{i,j}$ in (\ref{eq_diff_rate_1d}) by the overlap length relative to the size of the source subvolume, i.e.,
\begin{equation}
\label{eq_diff_rate}
\kth{i,j} = \frac{2\Dx{}h_o}{h_i^2(h_i + h_j)}.
\end{equation}

Eq.~(\ref{eq_diff_rate}) is consistent in two senses. First, if adjacent subvolumes have the \emph{same} molecular concentration (note that the concentration in subvolume $\subV{i}$ is $\molNum{i}/h_i^2$), then it can be shown that the propensities of molecules traveling in both directions will be \emph{equal} for any $h_o$, i.e., conservation laws are satisfied. Second, if multiple subvolumes collectively overlap a face of the source subvolume, as in example D) in Fig.~\ref{fig_sub_different}, then the sum of the transition rates into all of the destination subvolumes would not change if we merged those subvolumes. We also note that we can extend (\ref{eq_diff_rate}) to cubes in three dimensions \edit{by defining an overlap \emph{area} $A_o$ and scaling $\kth{i,j}$ by a factor of $A_o/h_i^2$ instead of $h_o/h_i$}{}.

One potential source of inaccuracy in applying (\ref{eq_diff_rate_1d}) is that, for a given number of molecules in $\subV{i}$, the outgoing transition rate will decrease as the destination $\subV{j}$ increases in size. The reason for this is that (\ref{eq_diff_rate_1d}) implicitly places the transition face at the midpoint of the centers of the two subvolumes, such that increasing the size of $\subV{j}$ also effectively increases the size of $\subV{i}$. We can mitigate the impact of this artefact by changing subvolume sizes \emph{gradually} over space.

\subsection{Transitions Across the Hybrid Interface}
\label{sec_hybrid_transition}

We now consider diffusion events across the interface between the microscopic regime $\Vol{\M}$ and the mesoscopic regime $\Vol{\subV{}}$, and vice versa. Molecules that are within the mesoscopic regime can only enter the microscopic regime if they transition from a subvolume that has a face shared with the boundary of the microscopic regime. Similarly, individual molecules that move outside the microscopic regime must be placed in a subvolume with a face along the interface with $\Vol{\M}$.

\begin{figure}[!tb]
	\centering
	\def\svgwidth{0.4\linewidth}
	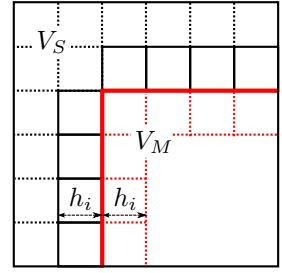
	\caption{Interface between the mesoscopic and microscopic regimes (shown as the red solid line). The solid lines form the subvolumes in $\Vol{\subV{}}$ that are along the interface. The dotted black lines form subvolumes that are not along the interface. The red dotted lines form ``virtual'' subvolumes that are within $\Vol{\M}$ and are used to initialize molecules that transition from $\Vol{\subV{}}$. Any transition between regimes is through a subvolume in $\Vol{\subV{}}$ that is along the interface.}
	\label{fig_hybrid_interface}
\end{figure}

To be effective at increasing computational efficiency, we expect that $\Vol{\M}$ would be smaller than $\Vol{\subV{}}$. Therefore, we restrict our partitioning such that a microscopic region must be ``surrounded'' by $\Vol{\subV{}}$ or the system boundary, as shown in Fig.~\ref{fig_hybrid_interface}. Any subvolume $\subV{}$ along the interface with $\Vol{\M}$ will have only one face along the interface between $\Vol{\subV{}}$ and the corresponding region in $\Vol{\M}$. Furthermore, we impose that all subvolumes that form an interface with a region in $\Vol{\M}$ should be the same size. This size restriction facilitates our implementation of transitions between regimes and is also consistent with our desire to limit simulation artefacts by changing subvolume sizes gradually over space.

For simplicity and clarity of exposition, we adopt the simplified transition rules described in \cite{RefWorks:870} (more complex rules such as those in \cite{RefWorks:869} will be considered in future work):

\emph{$\Vol{\subV{}}$ to $\Vol{\M}$}: To determine the transition rate from a subvolume along the interface to $\Vol{\M}$, we create a mirror ``virtual'' subvolume in $\Vol{\M}$ that is the same size, i.e., $h_j=h_i$, as shown in Fig.~\ref{fig_hybrid_interface}. If a diffusion event occurs such that a molecule must be added to $\Vol{\M}$, then we place the molecule at random within the corresponding ``virtual'' subvolume and it is then treated as an individual particle within $\Vol{\M}$.

\emph{$\Vol{\M}$ to $\Vol{\subV{}}$}: When individual molecules from $\Vol{\M}$ are identified to have entered $\Vol{\subV{}}$ after they are moved via (\ref{eq_diff_x}) and (\ref{eq_diff_y}), we calculate the distance from the molecules' new positions to the locations of the \emph{centres} of each of the subvolumes along the interface. For each individual molecule removed, we increment the molecule count of the subvolume whose centre is the closest. The corresponding propensities are updated and we re-generate the uniform random variables to determine the time and location of the next diffusion event in $\Vol{\subV{}}$.

\subsection{Hybrid Multi-Scale Simulation Algorithm}
\label{sec_alg}

We now describe our micro/meso simulation algorithm for hybrid multi-scale stochastic diffusion. The general structure of our algorithm is similar to the two-regime method in \cite{RefWorks:869}, with three significant differences. First, we apply simplified transition rules at the interface. Second, we specify that we partition the mesoscopic regime $\Vol{\subV{}}$ into subvolumes of different sizes; \cite{RefWorks:869} only considers subvolumes of equal size. Third, we describe the initialization and observation of molecules from the perspective of MC, i.e., the source is a transmitter (TX) and we observe the molecules at a receiver (RX).

We note that each regime has its own current time. The $\Vol{\M}$ time $t_{\M}$ is advanced in deterministic increments $\Delta t_{\M}$ and \emph{all} molecules in $\Vol{\M}$ are moved at the same time. The $\Vol{\subV{}}$ time $t_{\subV{}}$ is advanced based on a stochastic realization of the next diffusion event, when \emph{one} molecule is moved. We compare $t_{\M}$ and $t_{\subV{}}$ to determine the next regime to simulate. Thus, we generally update $t_{\subV{}}$ much more often than $t_{\M}$.

Realizations of the simulation algorithm execute as follows:
\begin{enumerate}
	\item \emph{Environment initialization}. Define the regimes $\Vol{\M}$ and $\Vol{\subV{}}$. $\Vol{\subV{}}$ is partitioned into subvolumes of different sizes as specified. Define $\Vol{\M}$ time step $\Delta t_{\M}$ and set $\Vol{\M}$ time $t_{\M}=\Delta t_{\M}$. Set global time $t=0$. Define final time $t_f$.
	\item \emph{Communication initialization}. Initialize molecules at the TX. If the TX is in $\Vol{\M}$, then define random coordinates within a specified ``virtual'' container for the individual molecules. If the TX is in $\Vol{\subV{}}$, then distribute the molecules over the corresponding set of subvolumes. Similarly, define the RX as a virtual ``container'' in $\Vol{\M}$ or a set of subvolumes in $\Vol{\subV{}}$. Define time between observations $t_{ob}$. Set observation index $i_{ob}=1$.
	\item \emph{Mesoscopic initialization}. If the TX is in $\Vol{\subV{}}$, then generate diffusion event time $t_{\subV{}}$ via (\ref{eq_meso_time}).
	\item \emph{Simulation control}. End if $t \ge t_f$. Otherwise, go to step 5) if $t_{\subV{}}\le t_{\M}$ or to step 6) if $t_{\subV{}}> t_{\M}$.
	\item \emph{Mesoscopic simulation}. Update $t = t_{\subV{}}$. Determine the source $\subV{i}$ and destination $\subV{j}$ of the diffusion event via \edit{weighted die rolls}{}. Update molecule counts $\molNum{i}$ and $\molNum{j}$. If the destination is within $\Vol{\M}$, then randomly place molecule within the corresponding ``virtual'' subvolume. Update reaction propensities and determine a new $t_{\subV{}}$ via (\ref{eq_meso_time}). Go to step 7).
	\item \emph{Microscopic simulation}. Update $t = t_{\M}$. Move all molecules in $\Vol{\M}$ via (\ref{eq_diff_x}) and (\ref{eq_diff_y}). If a molecule moved outside of $\Vol{}$, then reflect back inside. If molecules entered $\Vol{\subV{}}$, then place each in the closest subvolume along the interface, update propensities, and determine a new $t_{\subV{}}$ via (\ref{eq_meso_time}). Update $t_{\M}=t_{\M}+\Delta t_{\M}$.
	\item \emph{Observation at RX}. If $t \ge i_{ob}t_{ob}$, then record the number of molecules within the RX, $\molNum{RX}(t)$, and set $i_{ob} = i_{ob}+1$. Go to step 4).
\end{enumerate}

It is straightforward to extend this algorithm to simulate communication via multiple molecule releases by initializing more molecules at the TX when needed.

\section{Simulation Results}
\label{sec_results}

\begin{figure}[!tb]
	\centering
	\def\svgwidth{.8\linewidth}
	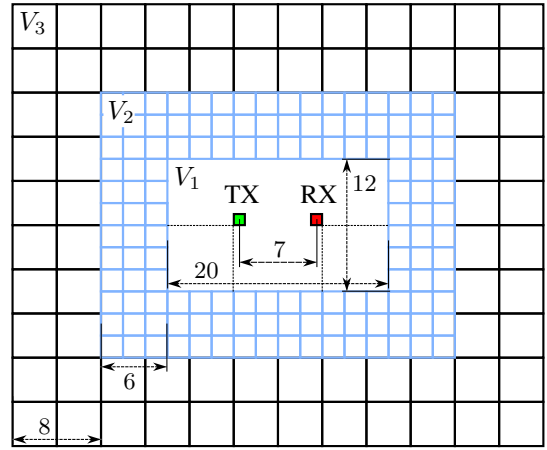
	\caption{Simulation environment drawn to scale. The transmitter (TX) and receiver (RX) regions are squares of width $h_1=1$. The inner region $\Vol{1}$ has width $20$ and height $12$. The middle region $\Vol{2}$ (in light blue) surrounds $\Vol{1}$ and has width $6$. The outer region $\Vol{3}$ (in black) surrounds $\Vol{2}$, has width $8$, and has a reflective outer boundary. The partitioning of $\Vol{}$ is shown for the multi-scale mesoscopic model (MESO-MS) in Table~\ref{table_partition}.}
	\label{fig_sim_environ}
\end{figure}

In this section, we present simulations to assess the accuracy and efficiency of multi-scale diffusion simulations in comparison with existing simulation models. We consider a sample dimensionless environment $\Vol{}$ of length $48$ and height $40$. The coefficient of diffusion is $\Dx{}=1$. We partition $\Vol{}$ into three regions, $\Vol{1}$, $\Vol{2}$, and $\Vol{3}$, as shown in Fig.~\ref{fig_sim_environ}. For a given model, each region is either in the microscopic or mesoscopic regime, as defined in Table~\ref{table_partition} \edit{(further study is required to more generally consider the effects of partitioning)}{}. When the $i$th region is mesoscopic, all subvolumes in that region have width $h_i$. The TX and RX are always squares of area 1, at the locations in $\Vol{1}$ shown in Fig.~\ref{fig_sim_environ}, but are in the same regime as the rest of $\Vol{1}$. Generally, we emphasize accuracy in $\Vol{1}$ and computational efficiency in $\Vol{3}$. As shown in Table~\ref{table_partition}, we consider a fully microscopic model (MICRO), a traditional mesoscopic model (MESO), a multi-scale mesoscopic model (MESO-MS), a regular hybrid model where the subvolumes are the same size (HYB), and a multi-scale hybrid model (HYB-MS), where the multi-scale models introduced in this paper have square subvolumes of different sizes. The total number of subvolumes in each model is also listed. The time step $\Delta t_{\M}$ used in all microscopic regions is $\Delta t_{\M}=0.25$.

\begin{table}[!tb]
	\centering
	\caption{System partitioning used for simulations. When a region uses the mesoscopic model, all subvolumes in that region have the width specified.}
	{\renewcommand{\arraystretch}{1.4}
		\begin{tabular}{|c|c|c|c|c|c|}
			\hline
			\bfseries Model & $\Vol{1}$ & $\Vol{2}$ & $\Vol{3}$ & $\#$ of $\subV{}$ & $\Delta t_{\M}$ \\ \hline
			MICRO & Micro & Micro & Micro & - & 0.25\\ \hline
			MESO & $h_1=1$ & $h_2=1$ & $h_3=1$ & 1920 & -\\ \hline
			MESO-MS & $h_1=1$ & $h_2=2$ & $h_3=4$ & 444 & - \\ \hline
			HYB & Micro & $h_2=1$ & $h_3=1$ & 1680 & 0.25\\ \hline
			HYB-MS & Micro & $h_2=1$ & $h_3=2$ & 816 & 0.25\\ \hline
		\end{tabular}
	}
	\label{table_partition}
\end{table}

\begin{figure}[!tb]
	\centering
	\includegraphics[width=\linewidth]{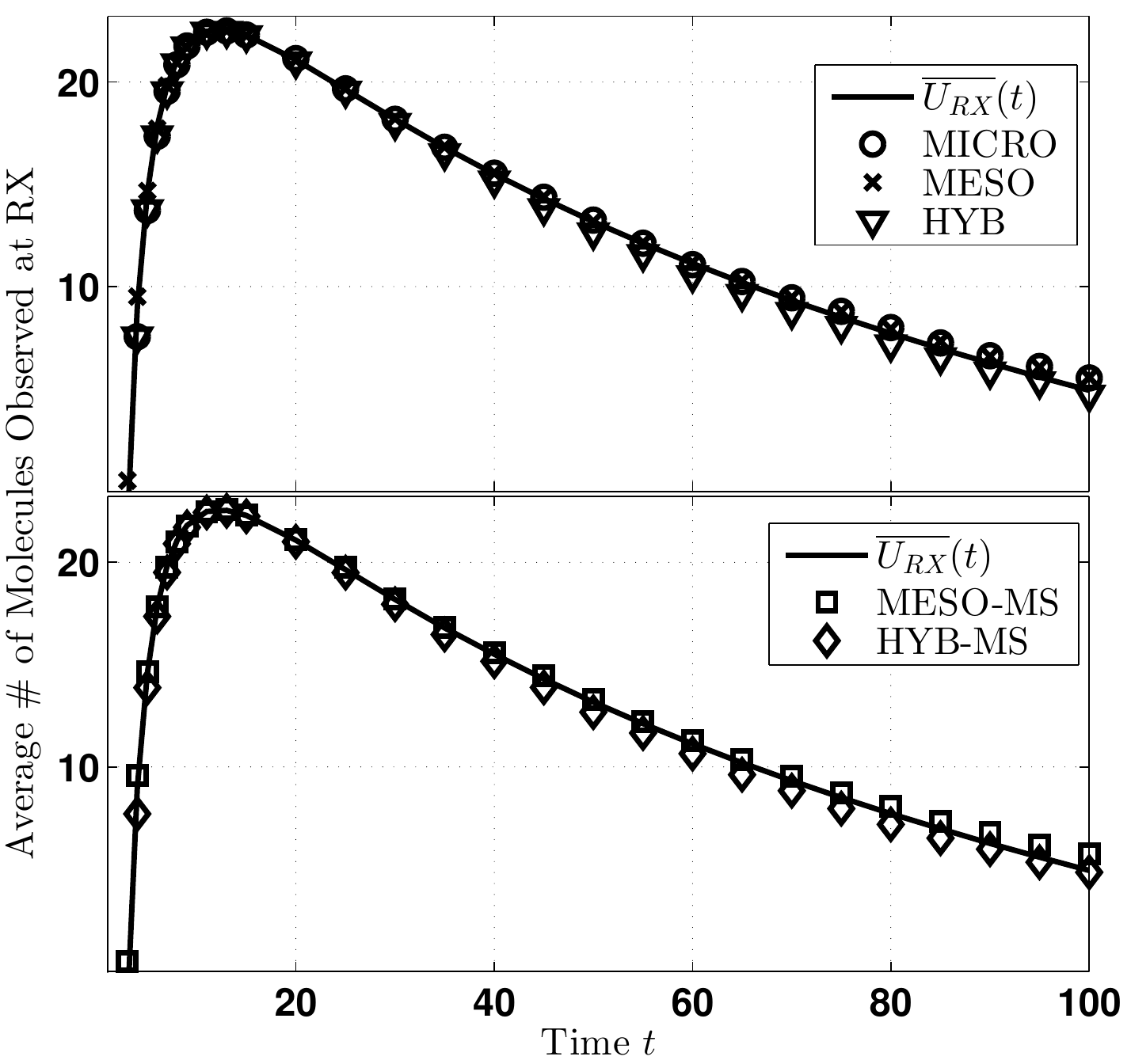}
	\caption{Impulsive release test, where molecules are initialized at the TX area in Fig.~\ref{fig_sim_environ} and the RX counts the number of molecules observed. All simulator models are compared with the number of molecules expected if the system were unbounded, i.e., $\overline{\molNum{RX}}(t)$ in (\ref{eq_mol_exp}), which is the same line in both plots. The existing models are compared in the upper plot and our multi-scale models are compared in the lower plot. Results are averaged over $10^4$ realizations and the vertical axes are logarithmic.}
	\label{fig_impulse}
\end{figure}

\begin{figure}[!tb]
	\centering
	\includegraphics[width=.9\linewidth]{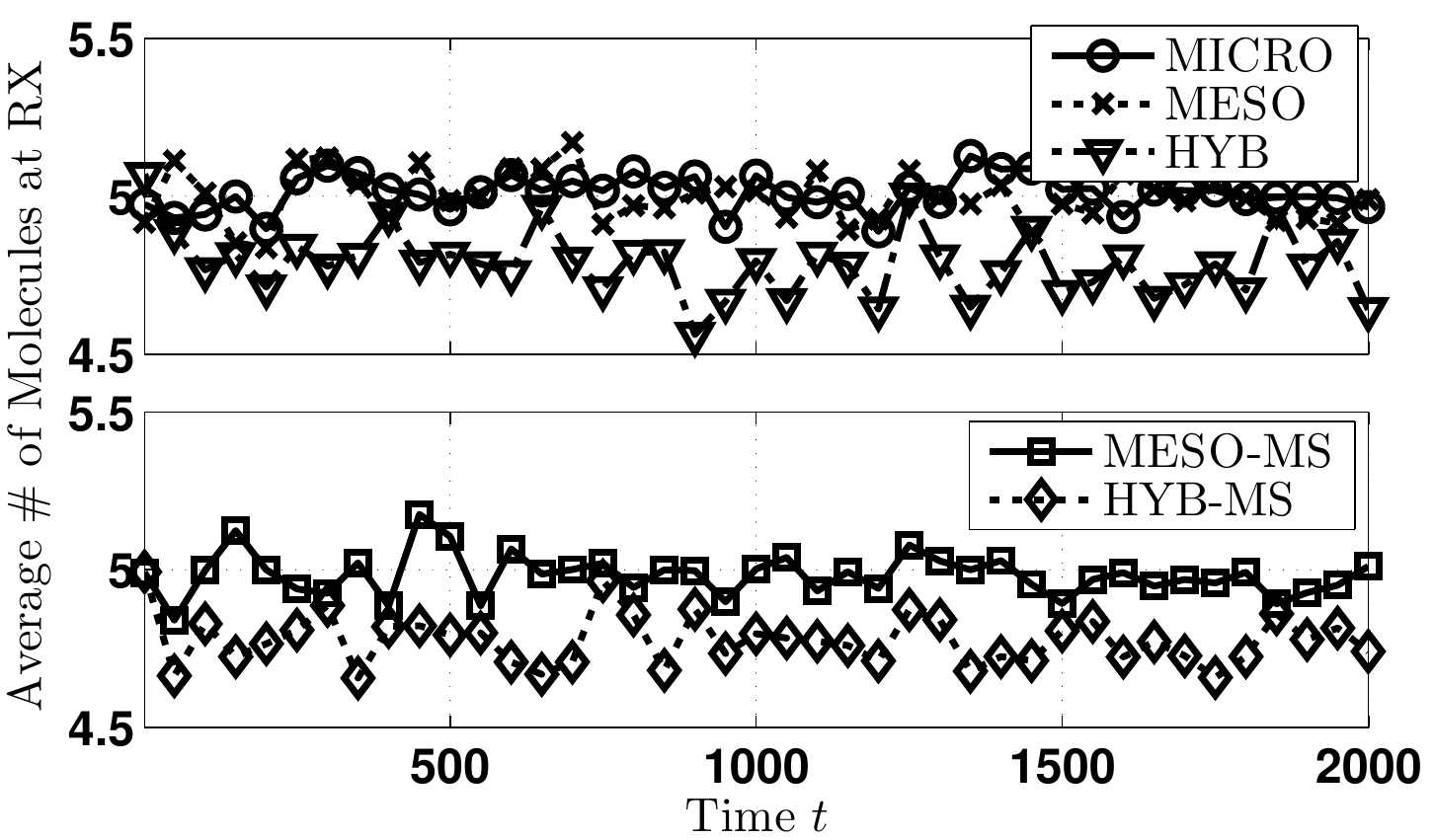}
	\caption{Uniform distribution test, where molecules are initialized over the entire environment in Fig.~\ref{fig_sim_environ} and the RX counts the number of molecules observed. The number of molecules expected at the RX is 5. The existing models are compared in the upper plot and our multi-scale models are compared in the lower plot. Results are averaged over $10^3$ realizations.}
	\label{fig_uniform}
\end{figure}

In the remainder of this section, we separately discuss the accuracy and the efficiency of the proposed models. We observe model accuracy in the impulsive release test in Fig.~\ref{fig_impulse} and in the uniform distribution test in Fig.~\ref{fig_uniform}. Then, we present the computational efficiency of all models in both tests in Table~\ref{table_complexity} by counting the number of \emph{molecule events} as a preliminary measure of computational complexity. We define two types of event: 1) updating the position of an individual molecule in $\Vol{\M}$ via (\ref{eq_diff_x}) and (\ref{eq_diff_y}), i.e., a ``Micro'' event; and 2) executing a diffusion event in $\Vol{\subV{}}$, i.e., a ``Meso'' event. If a molecule's position is updated in one regime and it must be transferred to the other regime, then we still count a single event. Having fewer events of a given type correlates to a faster execution time. A more formal assessment of the computational complexity of multi-scale simulation models is left for future work.

The \emph{impulsive release test} initializes molecules inside the TX area shown in Fig.~\ref{fig_sim_environ}. We place $10^4$ molecules at time $t=0$ and count the number of molecules at the RX until $t=100$. From \cite[Eq.~(3.4)]{RefWorks:586}, we can write the number of molecules expected at any point in an infinite plane due to an impulsive point source. If we assume that $\Vol{}$ is large enough to approximate as infinite, and if we assume that the sizes of the TX and the RX are small enough to approximate as points, then we can write the number of molecules expected at the receiver, $\overline{\molNum{RX}}(t)$, as
\begin{equation}
\label{eq_mol_exp}
\overline{\molNum{RX}}(t) = \frac{Nh_1^2}{4\pi\Dx{} t}\EXP{-\frac{d^2}{4\Dx{}t}},
\end{equation}
where $N$ is the number of molecules released, $h_1^2=1$ is the area of the RX, and $d=7$ is the distance between the centers of the TX and the RX. This approximation becomes less accurate as $t \to \infty$ and $\overline{\molNum{RX}}(t) \to 0$ because the simulated system is finite. Nevertheless, it is a useful benchmark for comparing the simulation models.

In Fig.~\ref{fig_impulse}, we plot the time-varying number of molecules observed at the RX for each simulator, averaged over $10^4$ realizations, and compare with the expected number of molecules $\overline{\molNum{RX}}(t)$. The MICRO and MESO models are very similar and very close to $\overline{\molNum{RX}}(t)$, showing that both can accurately model diffusion. $\overline{\molNum{RX}}(t)$ appears to be a good approximation and only starts to deviate from the MICRO and MESO models for $t > 75$. Our multi-scale MESO model, MESO-MS, appears to be just as accurate as the MESO model, even though we see from Table~\ref{table_partition} that the MESO-MS model uses significantly fewer subvolumes to represent the same system (444 versus 1920). The HYB model deviates slightly from the MICRO and MESO models for $t>20$ with fewer molecules observed, suggesting that the net transfer of molecules at the interface from $\Vol{1}$ to $\Vol{2}$ is too large. Nevertheless, our multi-scale hybrid model, HYB-MS, is comparably accurate with about half as many subvolumes representing the same system.

The \emph{uniform distribution test} assesses whether each model can maintain an initial uniform distribution of molecules. This test is particularly useful for the hybrid and multi-scale methods to ensure that there is no large net movement of molecules across the interface between regimes or between subvolumes of different sizes. We initialize by placing $9600$ molecules throughout $\Vol{}$, so that there is on average 5 molecules in every area of size 1. We then observe the number of molecules at the RX until $t=2000$ (i.e., $20$ times longer than the impulsive test), and plot the average over $10^3$ realizations in Fig.~\ref{fig_uniform}. For this test, the MESO-MS model appears to maintain the uniform distribution. The hybrid models appear to initially ``leak'' too many molecules into the mesoscopic regime, such that the mean number of observed molecules drops to about $4.75$, and this number is maintained for the remainder of the test. The initial leakage might be mitigated by the use of more accurate transition rules, such as those described in \cite{RefWorks:869}.

Finally, we compare model efficiency by counting the number of molecule events per molecule per unit time of each  model, averaged over all realizations, as shown in Table~\ref{table_complexity}. There are a number of meaningful comparisons to make, despite the simplified measure of complexity which does not account for the differences between microscopic and mesoscopic models. Each multi-scale model can be compared with its regular counterpart to observe the computational gains of representing mesoscopic regions with larger subvolumes outside the region of interest. We see that the MESO-MS model is much more efficient than the MESO model; the MESO-MS model is more than twice as fast for the impulsive test and more than four times as fast for the uniform test. The mesoscopic portion of the HYB-MS model is also significantly faster than that of the HYB model for both tests. The reason for the faster simulation of multi-scale models is that the total propensity of a region is smaller when the region is partitioned into larger subvolumes. This can be shown from (\ref{eq_diff_prop}), and the effect from (\ref{eq_meso_time}) is that diffusion events in regions with larger subvolumes occur less frequently. It is encouraging that the multi-scale models are faster but with negligible change in accuracy at the RX, especially given that the multi-scale models represent over half of the system (i.e., $\Vol{3}$) with subvolumes that are much larger than the size of the receiver (i.e., 16 times larger and 4 times larger for the MESO-MS and HYB-MS models, respectively). Future work will consider the degradation of accuracy at the RX as subvolume sizes in critical regions become too large.

\begin{table}[!tb]
	\centering
	\caption{Average number of microscopic and mesoscopic molecule events per molecule per realization per unit time, for the simulation tests presented in Figs.~\ref{fig_impulse} and \ref{fig_uniform}.}
	{\renewcommand{\arraystretch}{1.3}
		\begin{tabular}{|c|c|c|c|c|}
			\hline
			\multirow{2}{*}{\bfseries Model} & \multicolumn{2}{c|}{\bfseries Impulsive (Fig.~\ref{fig_impulse})} & \multicolumn{2}{c|}{\bfseries Uniform (Fig.~\ref{fig_uniform})} \\ \cline{2-5}
			& \bfseries Micro & \bfseries Meso & \bfseries Micro & \bfseries Meso \\ \hline
			MICRO & $4.00$ & - & $4.00$ & - \\ \hline
			MESO & - & $3.97$ & - & $3.91$ \\ \hline
			MESO-MS & - & $1.93$ & - & $0.892$ \\ \hline
			HYB & $1.50$ & $2.50$ & $0.48$ & $3.45$ \\ \hline
			HYB-MS & $1.50$ & $1.62$ & $0.48$ & $1.66$ \\ \hline
		\end{tabular}
	}
	\label{table_complexity}
\end{table}

We also note that the MESO and MESO-MS models clearly performed the uniform test faster than the impulsive test. One reason is that the uniform test places some molecules into subvolumes that are along the boundary of $\Vol{}$. If the $i$th subvolume is along the boundary of $\Vol{}$, then it has fewer neighbors and hence a smaller subvolume propensity $\prop{i}$. Thus, the total system propensity is lower and diffusion events initially occur less frequently than in the impulsive test (i.e., before the molecules in the impulsive test diffuse to a uniform distribution). Another advantage of multi-scale models in the uniform test is that the subvolumes near the boundary are larger, so diffusion events occur even less frequently.

\section{Conclusions}
\label{sec_concl}

In this paper, we presented the foundation of a multi-scale stochastic simulator for the study of diffusive MC. We combined a hybrid model with the use of subvolumes of different sizes in the mesoscopic regime. We showed that multi-scale variants \edit{can be}{} as accurate as existing models, despite using fewer subvolumes to describe the system. Multi-scale models increase computational efficiency by using larger subvolumes in regions far from the communication link.

Our on-going work is the development of a comprehensive simulator based on the foundation presented in this paper. We intend to include phenomena such as bulk fluid flow, boundary interactions, and chemical reaction pathways. We will extend to three dimensions with more accurate transition rules to mitigate molecule ``leakage'' between regimes. We are interested in strategies to provide more flexibility between accuracy and efficiency, such as ``tau-leaping'', where multiple events in the mesoscopic model are executed simultaneously; see \cite{RefWorks:612}. Further study also includes a more extensive comparison of the channel statistics when multi-scale approaches are used.

\bibliography{../references/nano_ref}

\end{document}